\documentclass{jetpl}

\usepackage[cp1251]{inputenc}
\usepackage[english,russian]{babel}

\twocolumn

\rus

\title{Хиральные оптические таммовские состояния
на границе среды с винтовой симметрией тензора диэлектрической проницаемости}
\rtitle{Хиральные оптические таммовские состояния \ldots}
\sodtitle{Хиральные оптические таммовские состояния
на границе среды с винтовой симметрией тензора диэлектрической проницаемости}

\author{И.\,В.\,Тимофеев\/\thanks{tiv@iph.krasn.ru}, С.\,Я.\,Ветров}
\rauthor{И.\,В.\,Тимофеев, С.\,Я.\,Ветров}
\sodauthor{Тимофеев, Ветров}

\address{Институт Физики им. Л.В. Киренского, Красноярский научный центр, СО РАН, Красноярск 660036, Россия}
\address{Лаборатория нелинейной оптики и спектроскопии, Сибирский Федеральный университет, Красноярск 660041, Россия}
\address{Институт инжинерной физики и радиоэлектроники, Сибирский Федеральный университет, Красноярск 660041, Россия}

\dates{28 июня 2016\,г.}{31 июля 2016\,г.}

\abstract{Аналитически и численно описано новое оптическое состояние
на границе хиральной среды, обладающей непрерывной винтовой симметрией тензора диэлектрической проницаемости. 
Рассмотрен случай, когда тангенциальное волновое число равно нулю. Локализуемое вблизи границы состояние не переносит энергии вдоль этой границы и экспоненциально спадает по мере удаления от границы.
Проникновение поля вглубь хиральной среды блокируется на длинах волн, соответствующих фотонной запрещенной зоне и близких к шагу винта. 
При этом поляризация света вблизи границы имеет тот же знак хиральности, что и винтовая симметрия.
Показано, что однородная окружающая среда, либо подложка должна проявлять анизотропное отражение металлического типа.
Спектральное проявление состояния определяется углом между оптическими осями сред на границе.
В качестве конкретного примера рассмотрено состояние на границе холестерического жидкого кристалла и анизотропного металл-диэлектрического нанокомпозита.}


\begin{document}

\maketitle 

Поверхностное состояние, которое можно наблюдать при падении света по нормали к поверхности, в оптике называют оптическим таммовским состоянием (ОТС) \cite{Vinogradov2010r}. 
Локализованный свет можно представлять запертым вблизи общей границы двух зеркал, понимая под зеркалами среды, на границе которых происходит отражение света.
В литературе известны различные типы зеркал, способных отражать свет при нормальном падении. К ним относятся зеркала металлического и фотонно-кристаллического типа.

Особо выделяется фотонно-кристаллическое отражение в средах, не обладающих зеркальной симметрией оптических свойств, но обладающих непрерывной винтовой симметрией тензора диэлектрической проницаемости. Будем называть такие среды \emph{хиральными}. К ним относится холестерический жидкий кристалл (холестерик), состоящий из ориентированных молекул, преимущественное направление которых закручивается в пространстве в виде винтовой спирали \cite{Belyakov1992b}. Другим примером может служить закрученная наклонная скульптурированная тонкая пленка \cite{LakhtakiaMessier200501book}.
Винтовая периодичность приводит к дифракции (объемному отражению) лишь света, поляризованного по кругу в направлении, соответствующем закручиванию винтовой спирали.
В отличие от нехиральных (зеркально-симметричных) фотонных кристаллов свет с противоположно закрученной круговой поляризацией не дифрагирует.

ОТС были обнаружены как на границе двух нехиральных сред \cite{Vinogradov2010r}, так и на границе двух хиральных зеркал \cite{Schmidtke2003_EPJ} -- в виде дефектных мод.
Однако, по нашим сведениям, до сих пор не удавалось получить ОТС
на границе хирального и нехирального зеркал.
Сложность возникает из-за того, что изотропное зеркало меняет поляризацию света и дифрагирующая волна хирального фотонного кристалла перетекает в недифрагирующую. В результате волна испытывает не более двух циклов отражений, после чего покидает границу зеркала \cite{Timofeev2013}.
В данной работе представлена попытка избавиться от описанного затруднения при помощи анизотропной подложки. Рассмотрен случай нормального падения, когда перенос энергии вдоль поверхности отсутствует и волновой вектор не имеет касательной составляющей.

Представим анизотропную подложку как одноосный кристалл,
оптическая ось которого направлена вдоль оси $x$, а сама среда расположена в полупространстве $z<0$.
Поле в нем раскладывается на необыкновенную и обыкновенную волны.
Уравнение для волн, распространяющихся в направлении противоположном направлению оси $z$, имеет следующий вид:
	\begin{align}
	\label{eq:NC} \nonumber
	\left[ {\begin{array}{r} 
	E_x \\ H_y \\ E_y \\ -H_x 
	\end{array}} \right] =
	E_x^0 \left[ {\begin{array}{c}
	1 \\ -n^0_e \\ 0 \\ 0
	\end{array}} \right]
	\exp \left( {-i \kappa n^0_e z - i \omega t} \right) 
	\\ 
	+E_y^0 \left[ {\begin{array}{c}
	0 \\ 0 \\ 1 \\ -n^0_o
	\end{array}} \right]
	\exp \left( {-i \kappa n^0_o z - i \omega t} \right),
	\end{align}
где
$E_{x,y},H_{x,y}$ – комплексные проекции напряженностей электрического и магнитного полей. 
$\kappa = \omega/c$ – волновой вектор в вакууме.
$n^0_{e,o}$ – необыкновенный и обыкновенный показатели преломления (ПП) подложки.

Для хиральной среды собственная волна может быть описана следующим образом \cite{Belyakov1992b}:
	\begin{align}
	\label{eq:ChLC} 
	& \left[ {\begin{array}{r} \nonumber
	E_x \\ H_y \\ E_y \\ -H_x 
	\end{array}} \right] = &\\
	 &A\left[ {\begin{array}{c}
	1 \\ (q + \tau)/\kappa \\ -i \\ -i (q + \tau)/\kappa
	\end{array}} \right]&
	\exp \left( {i\left( {q z+\tilde{\varphi}(z) - \omega t} \right)} \right) \nonumber \\ 
	+&B\left[ {\begin{array}{c}
	1 \\ - (q - \tau)/\kappa \\ i \\ -i (q - \tau)/\kappa \\ 
	\end{array}} \right]&
	\exp \left( {i\left( {q z- \tilde{\varphi}(z) - \omega t} 
	\right)} \right).
	\end{align}
Здесь $A,B$ -- комплексные амплитуды волн, поляризованных по кругу вдоль винта и бегущих в противоположных направлениях, вперед (Ahead) вдоль оси z и назад (Back). 
Угол закручивания оптической оси $\tilde{\varphi}(z) = \tau z+\varphi$ отсчитывается от оси $x$ в направлении оси $y$;
$\tau= 2\pi/p$ – волновой вектор закручивания оптической оси;
$p$ – шаг винтовой спирали;
волновой вектор $q$ имеет значение
\[
q = \sqrt{\tau^2+\epsilon \kappa^2-2 \tau \kappa \sqrt{\epsilon+\delta^2 \kappa^2/4 \tau^2}}.
\]
Компоненты тензора диэлектрической проницаемости $\epsilon_{\parallel,\perp} = \epsilon \pm \delta$.

Для дифрагирующей волны волновой вектор $q$ принимает чисто мнимое значение в запрещенной зоне частот, выражаемой неравенством
	\begin{equation}
		\label{eq:ChLC_PBG}
	\frac{\tau}{\sqrt{\epsilon+\delta}} <
	\kappa = \frac{\omega}{c} 
	< \frac{\tau}{\sqrt{\epsilon-\delta}}.
	\end{equation}
В запрещенной зоне напряженности волн $A$ и $B$ представляют собой вектора одинаковой длины, а их разность фаз $\Phi$ зависит от частоты и изменяется от $0$ до $\pi$:
	\begin{align} 
	\frac{A}{B} = e^{+i\Phi(\kappa)} 
	= \frac{(q-\tau)^2/\kappa^2-\epsilon}{\delta}, \nonumber
\\
	\nonumber
	\frac{B}{A} = e^{-i\Phi(\kappa)} 
	= \frac{(q+\tau)^2/\kappa^2-\epsilon}{\delta}.
	\end{align}
Для дальнейшего удобно избавиться от $q$, переписав уравнения в виде
	\begin{align} \nonumber
	{\kappa}{\sqrt{\epsilon+\delta exp(-  i \Phi(\kappa))})} = {\tau - q},
	\\  \nonumber
	{\kappa}{\sqrt{\epsilon+\delta exp(+ i \Phi(\kappa))})} = {\tau + q},
	\end{align}
и почленно сложив получившиеся равенства:
	\begin{align} 
		\label{eq:AB}
	{\kappa}{Re(\sqrt{\epsilon+\delta exp(i \Phi(\kappa))})} = {\tau}.
	\end{align}

Удобно нормировать все волновые вектора и ПП на средний ПП хиральной среды,
тогда ПП подложки:
	\begin{equation} 
	n_{e,o} = n^0_{e,o}/\sqrt{\epsilon}.
	\end{equation}
Также использовано приближение малой анизотропии $\delta \ll \epsilon$.
При этом отношение амплитуд электрических и магнитных напряженностей приблизительно равно единице:
$(q \pm \tau)/\kappa \approx \pm 1$.
Условие сшивки полей на границе может быть получено непосредственным приравниванием напряженностей (\ref{eq:NC},\ref{eq:ChLC}) на границе $z=0$ в момент времени $t=0$:
	\begin{align}
	\label{eq:Boundary4D}
	&E_x^\varphi \left[ {\begin{array}{c}
	1 \\ -n_e \\ 0 \\ 0
	\end{array}} \right]
	+E_y^\varphi \left[ {\begin{array}{c}
	0 \\ 0 \\ 1 \\ -n_o
	\end{array}} \right]
	=
	r\left[ {\begin{array}{r}
	1 \\ 1 \\ -i \\ -i
	\end{array}} \right]
	+\left[ {\begin{array}{r}
	1 \\ -1 \\ i \\ -i
	\end{array}} \right],
	\\
	\label{eq:AmpReflection}
	&r = exp(i\Phi(\kappa) + {2i\varphi} ).
	\end{align}
Здесь $E_{x,y}^\varphi = E_{x,y}^0/B \exp ( {i\varphi} )$.
Четыре неизвестных для подложки выражаются через $r$ как
	\begin{align} 
	E_{x}^\varphi = 1 + r; 
	E_{y}^\varphi = i (1 - r); \nonumber 
	\\ 
	n_e = \frac{1-r}{1 + r};
	n_o = \frac{1 + r}{1 - r}. \label{eq:Boundary}
	\end{align}
По физическому смыслу $r$ -- это амплитудный коэффициент отражения от подложки.
Выражая $r$ через каждый из ПП, приходим к известным уравнениям Френеля для обыкновенной и необыкновенной волн:
	\begin{align}
	r_e =  r = \frac{1-n_e}{1+n_e};
	r_o = -r = \frac{1-n_o}{1+n_o}.	\label{eq:Fresnel}
	\end{align}
Одна из этих волн отражается в противофазе. Следовательно дифрагирующая круговая волна отражается опять в дифрагирующую.
Вообще, подложка, удовлетворяющая условию инверсии ПП $n_e = 1/n_o$, представляет собой сохраняющее поляризацию анизотропное зеркало \cite{Rudakova2016rm}.

Найденное решение (\ref{eq:Boundary}) обеспечивает фазовое согласование. После отражения от двух зеркал волна должна вернуться в прежнее состояние в той же фазе, обеспечивая конструктивную интерференцию и резонанс. Сдвиг частоты компенсирует угол $\varphi$ между оптическими осями на границе, поскольку для света круговой поляризации угол пространственного поворота вокруг направления распространения количественно равен фазе волны. 
Плавное изменение этого угла на $\pi$ при двукратном отражении изменяет фазу на $2\pi$ и позволяет добиться фазового согласования на границе зеркал.

Частоту ОТС можно выразить из второго уравнения сшивки (\ref{eq:AmpReflection}), отвечающего за согласование фазы, переписав его в виде
	\begin{align} \nonumber
	\Phi(\kappa) = \rho - {2\varphi},
	\end{align}
где $\rho$ -- комплексная фаза амплитудного коэффициента отражения $r= | r | exp(i\rho)$, определенная с точностью до $2\pi$.
С использованием выражения (\ref{eq:AB}) для набега фазы $\Phi(\kappa)$ в холестерике это дает явную зависимость частоты ОТС от угла $\varphi$:
	\begin{align} 
	\label{eq:GPhase}
	{\kappa} = \frac{\omega}{c} = \frac{\tau}{Re(\sqrt{\epsilon+\delta exp(i \rho -2i \varphi)})}.
	\end{align}
Соотношение (\ref{eq:GPhase}) определяет спектральное проявление ОТС внутри запрещенной зоны (\ref{eq:ChLC_PBG}) хиральной среды. Оно выполняет роль дисперсионного соотношения, поскольку тангенциальный волновой вектор описываемого ОТС равен нулю и не может определять частоту состояния. Решение при ненулевом тангенциальном волновом векторе выходит за рамки данной статьи.

Еще одно условие существования ОТС -- \emph{локализация поля} вблизи границы.
Со стороны хиральной среды локализация обеспечивается мнимым значением волнового вектора $q$ в запрещенной зоне.
Со стороны подложки для локализации требуется, чтобы оба ПП имели положительную мнимую часть, которая соответствует затуханию при выбранных знаках комплексного множителя $exp(ikz-i\omega t)$.
Однако в рассмотренном приближении из условия инверсии ПП $n_e = 1/n_o$ вытекает: $Im(n_e) \cdot Im(n_o) < 0$.
Поэтому для совместимости соотношений (\ref{eq:Boundary}) с локализацией корректно говорить лишь о предельном случае малой мнимой части, когда добротность состояния и длина локализации одновременно стремятся к бесконечности.
Конечные значения мнимой части ПП подложки приводят к потерям через недифрагирующую поляризацию и через поглощение металлического типа.

Построим локализованное состояние численным методом Берремана.
В качестве хиральной среды рассмотрим правозакрученный холестерик с анизотропией $\delta = 0.2$ и нормированным шагом винтовой спирали $p \sqrt{\epsilon}=500$~нм, толщина слоя составляет пять шагов винтовой спирали.
В качестве подложки рассмотрим нанокомпозит серебряных сфероидов, сплющенных в направлении $x$ и помещенных в матрицу с ПП равным среднему ПП холестерика.
Использование формулы Максвелл-Гарнетта позволяет подобрать параметры нанокомпозита так, что $n_o = (1+i)n_m$; $n_e = (1+i)/2n_m$ на определенной частоте в видимом диапазоне длин волн \cite{Rudakova2016rm}.
Положим $n_m = 10$.
Дисперсия нанокомпозита не учитывается.
Справедливо условие $n_e = 1/n_o^*$, которое подменяет условие инверсии ПП, вытекающее из аналитического решения (\ref{eq:Boundary}).
Звездочка $^*$ означает комплексное сопряжение.
Предлагаемая подмена обеспечивает соотношение на амплитудные коэффициенты отражения подложки $r_e = -r_o^*$, $Re(r_e) = -Re(r_o)$. В результате отражение в дифрагирующую поляризацию согласуется с условиями (\ref{eq:AmpReflection}) и (\ref{eq:GPhase}). Причем фаза отражения $\rho = 0$, так как амплитуда отражения имеет положительное действительное значение. При $sin(\varphi)<0$ ОТС не реализуется.

\begin{figure}[htbp]
\includegraphics{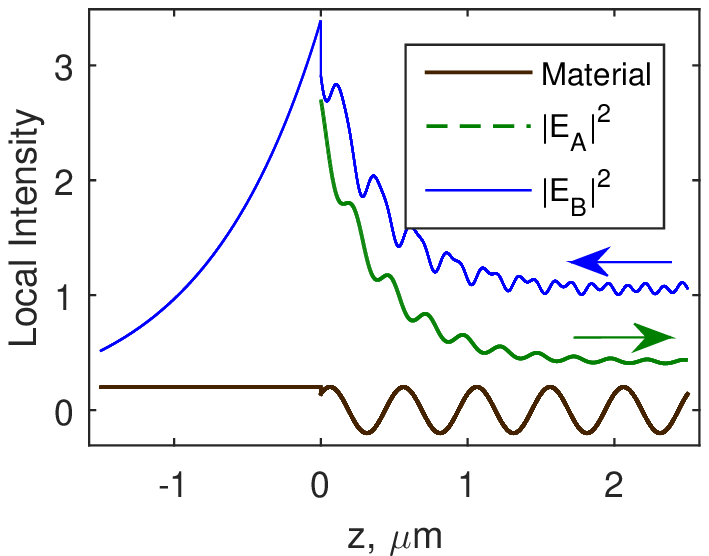}
\includegraphics{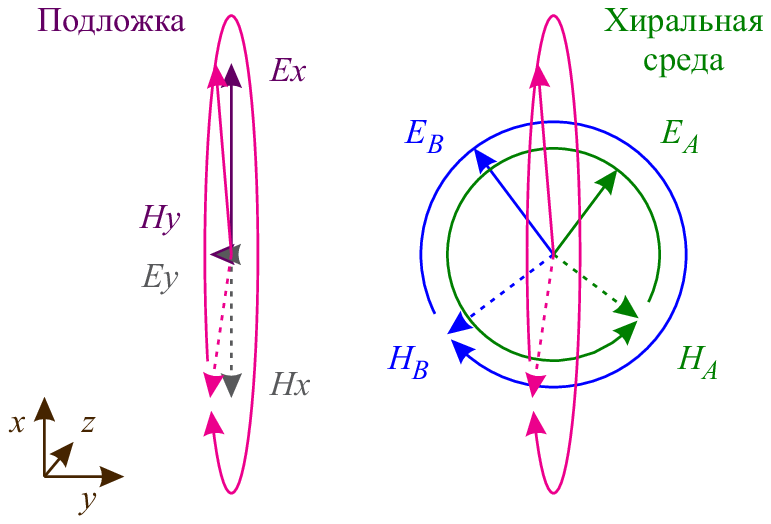}
\caption{
Рис.~1. а) Локальная интенсивность поля в зависимости от расстояния до границы.
`Material' -- схема ориентации оптической оси в подложке (прямая) и хиральной среде (синусоидальная проекция на ось $x$).
б) Сшивка полей на границе. Сплошными стрелками обозначены вектора напряженности электрического поля, штриховыми -- магнитного. В хиральной среде поле представляется в виде падающей на границу ($B$) и отраженной от нее ($A$) круговых волн. В подложке поле раскладывается на необыкновенную ($x$) и обыкновенную ($y$) волны.}
\label{fig:CSS_Boundary_continuity}
\end{figure}

На рис. \ref{fig:CSS_Boundary_continuity},а 
изображено ОТС в виде локальной интенсивности (квадрат амплитуды напряженности электрического поля) в зависимости от расстояния до границы. Локальная интенсивность нормирована на интенсивность волны, возбуждающей ОТС, падающей справа, из хиральной среды, и имеющей левую круговую поляризацию. Представлен результат прямого численного расчета методом Берремана.
Волна $B$, распространяющаяся влево, включает в себя волну, возбуждающую ОТС, и поэтому имеет б\'{о}льшую интенсивность, чем переотраженная от подложки волна $A$,
распространяющаяся вправо и отмеченная пунктиром.
Чтобы не загромождать рисунок, не показана суммарная локальная интенсивность $|A+B|^2$, которая вблизи границы почти в 7 раз превышает локальную интенсивность волны, возбуждающей ОТС. Также в выбранном масштабе не виден вклад обыкновенной волны в подложке, так как эта волна быстро затухает с глубиной подложки и имеет малую амплитуду электрической напряженности.
Чтобы получить гладкую экспоненциально спадающую огибающую, согласующуюся с уравнениями (\ref{eq:NC},\ref{eq:ChLC}), следует вычесть из решения волну, возбуждающую ОТС. 

На рис. \ref{fig:CSS_Boundary_continuity},б 
поясняется условие сшивки полей на границе. Эллипс поляризации результирующего поля вытянут в направлении $x$ как для электрической, так и для магнитной напряженностей. В хиральной среде его большая полуось пропорциональна сумме амплитуд $|A|+|B|$, а малая -- их разности $|A|-|B|$. В подложке для необыкновенной волны вытянута электрическая напряженность, а для обыкновенной -- магнитная. Равенство отношений большой и малой полуоси эллипса в хиральной среде и подложке дает:
	\begin{equation}
	\label{eq:ABBoundary}
	\frac{|A|+|B|}{|A|-|B|} =
	\frac{1}{n_e} =
	n_o \gg 1.
	\end{equation}
Это согласуется с аналитическим решением (\ref{eq:Boundary}).

\begin{figure}[htbp]
\includegraphics{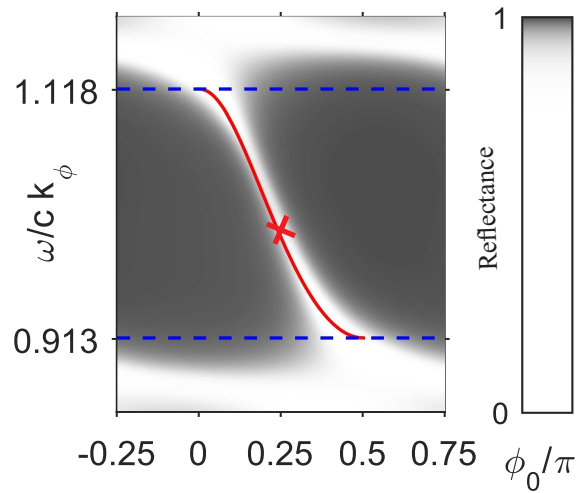}
\caption{
Рис.~2. 
Спектр отражения границы при различных углах $\varphi$ между оптическими осями на границе. 
Синим пунктиром обозначены края фотонной запрещенной зоны. 
Красная сплошная кривая отражает аналитическую зависимость (\ref{eq:GPhase})
Красный крестик при $\varphi=\pi/4$ и $\kappa=\tau$ соответствует параметрам рис.~\ref{fig:CSS_Boundary_continuity},а.
}
\label{fig:CSS_Spectr}
\end{figure}

На рис.~\ref{fig:CSS_Spectr} приведен спектр отражения от границы для света правой круговой поляризации, падающего из правозакрученной хиральной среды перпендикулярно границе.
Зависимость (\ref{eq:GPhase}) частоты ОТС от угла $\varphi$ качественно соответствует аналогичной зависимости  от фазы модуляции в нехиральном фотонном кристалле (см. \cite{Belyakov1992_ShilinaScalar} и ссылки в ней).

Плавное вращение зеркал приводит к смещению частоты ОТС вплоть до краев запрещенной зоны (\ref{eq:ChLC_PBG}). 
При совпадении оптических осей ОТС находится на высокочастотном краю зоны.
Середине зоны соответствует наиболее сильный провал в отражении при угле $\varphi=\pi/4$.
При возбуждении светом правой круговой поляризации отражение в провале составляет 90\% (красный крестик на рис.~\ref{fig:CSS_Spectr}), а для света левой круговой поляризации -- 45\% (рис.~\ref{fig:CSS_Boundary_continuity},а).
Величина угла $\varphi=\pi/4$ объясняется тем, что наибольший градиент показателя преломления хиральной среды наблюдается под углом $\pi/4$ к оптической оси, и электрическое поле, ориентированное в этом направлении, испытывает сильное объемное отражение.
ОТС уходит на низкочастотный край зоны при перпендикулярных оптических осях.
Для углов, больших $\pi/2$, провал отражения отсутствует,
так как не выполняется условие (\ref{eq:GPhase}).
\vspace{1em}


Аналитически решена задача о наличии ОТС на границе хирального и нехирального зеркал при нулевом тангенциальном волновом векторе. Решение накладывает сильное условие на параметры однородной подложки, на границе которой с хиральным зеркалом возможно хиральное ОТС.
Добротность состояния стремится к бесконечности лишь в случае длины локализации, стремящейся к бесконечности.
Найденное ОТС локализовано вблизи границы и экспоненциально спадает в обе стороны.
При проходе границы фаза регулируется вращением зеркал в плоскости границы, что описывается условием (\ref{eq:GPhase}), выполняющим роль дисперсионного соотношения. Аналитическая зависимость согласуется с результатами прямого численного расчета.

\clearpage


\end{document}